\documentclass[11pt]{article}

\usepackage{preamble}
\usepackage{multicol}
\usepackage{lineno}

\usepackage[firstpage=true]{background}
\backgroundsetup{contents={\parbox{6.5in}{\snowmass}}, scale=1,placement=top,opacity=1,color=black,position={3.25in,1.2in}}

\title{Innovations in trigger and data acquisition systems for next-generation physics facilities \\
\vspace{1em}
\Large A Snowmass 2022 White Paper}

\author[1]{Rainer Bartoldus\thanks{Rainer.Bartoldus@slac.stanford.edu}}
\author[1]{Catrin Bernius\thanks{Catrin.Bernius@slac.stanford.edu}}
\author[2]{David W. Miller\thanks{David.W.Miller@uchicago.edu}}

\affil[1]{SLAC National Accelerator Laboratory}
\affil[2]{Department of Physics, University of Chicago}

\begin{document}

\maketitle

\begin{abstract}
Data-intensive physics facilities are increasingly reliant on heterogeneous and large-scale data processing and computational systems in order to collect, distribute, process, filter, and analyze the ever increasing huge volumes of data being collected. 
Moreover, these tasks are often performed in hard real-time or quasi real-time processing pipelines that place extreme constraints on various parameters and design choices for those systems. 
Consequently, a large number and variety of challenges are faced to design, construct, and operate such facilities. 
This is especially true at the energy and intensity frontiers of particle physics where bandwidths of raw data can exceed 100 TB/s of heterogeneous, high-dimensional data sourced from 300M+ individual sensors. 
Data filtering and compression algorithms deployed at these facilities often operate at the level of 1 part in $10^5$, and once executed, these algorithms drive the data curation process, further highlighting the critical roles that these systems have in the physics impact of those endeavors. 
This White Paper aims to highlight the challenges that these facilities face in the design of the trigger and data acquisition instrumentation and systems, as well as in their installation, commissioning, integration and operation, and in building the domain knowledge and technical expertise required to do so. 
\end{abstract}

\clearpage

{
 \hypersetup{linkcolor=black}
 \setcounter{tocdepth}{1}
 \tableofcontents
}

\section{Executive Summary}
\label{sec:executive}

Trigger and Data AcQuisition (TDAQ) systems form a critical interface between detector instrumentation and computing systems at current and future physics facilities, and are comprised of a wide range and variety of devices. Capabilities and performance of the TDAQ systems deployed by experiments at these facilities have a direct impact on the amount and quality of the data and the physics that can be extracted from them. Innovative solutions to the challenges that these systems will face are essential to perform discovery science far into the future. Three categories of challenges are identified and discussed as being priorities for current and future research and development for TDAQ systems:
\begin{itemize}
    \item Future TDAQ systems will have to leverage currently existing and available hardware, firmware, and software in new ways.
    \item TDAQ developers will have to find ways to adopt and incorporate the unique technical requirements of physics facilities into new hardware concepts.
    \item The TDAQ community must ensure that the people, knowledge, and experience required to build, commission, and operate new TDAQ systems are fully supported and retained in the field.
\end{itemize}

\clearpage

\section{Introduction}
\label{sec:introduction}

As detector instrumentation technologies for signal generation advance and proliferate -- from novel microscopic quantum sensors to colossal integrated wire chambers -- the requirements and challenges to acquiring, processing, distributing, and interpreting those signals, often in real-time, also increase and intensify. Moreover, as the physics questions addressed by these facilities grow both more sophisticated and more subtle, the need to collect more data, faster, and with higher fidelity also grows extremely rapidly.

In this White Paper, we summarize several of the ``grand challenges'' that we envision may or must be met by innovations in Trigger and Data AcQuisition (TDAQ) systems for next-generation physics facilities. These challenges will then be discussed in the subsequent sections, with examples of existing or proposed solutions, drawing both on the Snowmass Community letters of interest as well as current literature in the field.

We identify three overarching themes that characterize both the challenges and the opportunities that will shape the future of innovation in TDAQ:
%
\begin{itemize}
    \item Novel applications and uses of current generations of commodity and custom hardware, firmware, and software for TDAQ systems;
    \item Confronting state-of-the-art devices and TDAQ paradigms with the specialized needs of future low-latency, high-throughput, and high-performance physics facilities;
    \item Building, integrating, commissioning, and operating large-scale, heterogeneous, and dynamic TDAQ systems at future facilities, including the acquisition and retention of the required domain knowledge and technical expertise.
\end{itemize}

These themes will frame the discussion in each of the sections below, which focus on the challenges faced by next-generation physics facilities, as well as a range of concepts and proposals for meeting these challenges with innovations in TDAQ devices and systems. 
\Secref{challenges} discusses sets of known obstacles facing TDAQ systems at current and currently-envisioned physics facilities and experiments. \Secref{heterogeneous} focuses on the trend towards heterogeneous TDAQ systems and the difficulties and opportunities that this presents both in the design of such a system as well as in its operation and interpretation. \Secref{systems} emphasizes the unique complexities of and potential solutions to the extreme signal processing environments of future facilities and experiments with dramatically higher numbers of sensors and data acquisition and transmission channels. Lastly, although it is also discussed in the earlier sections, \Secref{operation} and \Secref{people} concentrate on the distinct challenge of bringing a novel idea to life by establishing, growing, and maintaining a technical knowledge base amongst both physicists and technical staff for integrating, operating, and maintaining novel TDAQ systems over the lifetime of an experimental physics facility.

\comment{

\begin{itemize}
    \item Designing, building, integrating, and operating large-scale, heterogeneous, and dynamic physics facilities for data acquisition, filtering, processing, and storage using both commodity and custom hardware, firmware, and software.
    \begin{itemize}
        \item What are the particular challenges that we face that are niche or nonexistent in industrial applications? Exceptionally large numbers of devices/channels; data structures much more well-defined; temporal structures of the data; high-rate applications; high bandwidth requirements; data integrity and robustness; reproducibility
        \item How can we take devices, firmware, and software designed by industry for industrial applications and build the systems we need to do physics with them? What innovations are needed to do this?
        \item What is the current state of the art for heterogeneous digital and analog solutions? Dynamic and reconfigurable hardware? Deployment of advanced algorithms (AI/ML)? heterogeneous CPU/GPU/FPGA/SoC systems?
        \item What innovations are being considered or needed for achieving this?
        \item What physics selection and filtering algorithms and architectures can be deployed on these commodity systems? What are the advantages and potential physics impacts of that approach above and beyond the current state of the art?
    \end{itemize}
    \item Pushing the envelope of industry standard devices and systems, and their intended goals, by confronting these commodity solutions with the specific and specialized needs of low-latency, high throughput and high performance physics facilities. Considering and exploring the paradigm shifting approaches that can take TDAQ systems to new levels, taking advantage of new technologies and opportunities in new ways.
    \begin{itemize}
        \item What can the scientific world feed back into the industrial world in order to adapt technologies and/or create new ones that meet the goals and challenges of next-generation physics facilities?
        \item Further integration of AI engines into devices; wireless 5G command and control; 
        \item What innovations are being considered or needed for achieving this?
        \item Explore clock-driven vs. event-driven approaches and innovations, synchronous vs. asynchronous trigger systems, and the traditional boundary between them
        \item Where are we going with the next-generation of high-level synthesis? Will the latency and resource requirements of applications for reconfigurable devices be able to be met by the optimization procedures of future HLS or HLS-like software/firmware? What will HLS be like in 20 years?
    \end{itemize}
    \item Installation, integration, validation and operation of the developed systems
    \begin{itemize}
        \item What foundational and operational skills and training do we need to be able to achieve the goals?
        \item What are the interfaces to other areas, fields etc?
        \item Acquisition of domain knowledge
        \item Recruit talent to do the actual work
        \item Expertise that we need to build
        \item Connect to the design, building, and execution
        \item Need pool of people who actually know about devices, etc.
        \item Workforce development 
    \end{itemize}
\end{itemize}

}

\section{Challenges faced by current and future physics facilities}
\label{sec:challenges}

Current and currently-envisioned physics facilities face a variety of obstacles to the efficient acquisition, processing, and distribution of data throughout the system. 
In the energy and intensity frontiers of particle physics, as accelerator operating conditions at particle colliders evolve towards higher energies, intensities, and data rates, the signatures of interest become more and more challenging to analyze and select in the high particle density environment. As experimental designs expand both fiducial detector volumes and channel multiplicities by orders of magnitude, timely processing and interpretation become extraordinarily challenging. 

The current state of the art for heterogeneous digital and analog solutions needs to be pushed further to allow for dynamic and reconfigurable hardware, the usage of heterogeneous systems (e.g. Central Processing Units (CPUs), Graphical Processing Units (GPUs), Field Programmable Gate Arrays (FPGAs), Application Specific Integrated Circuits (ASICs), and Systems-on-Chip (SoCs)) and the possibility to deploy advanced algorithms (Artificial Intelligence (AI), Machine Learning (ML)~\cite{FastMLWhitePaper}) on the chosen configuration (see also \Secref{heterogeneous}).
Following the choice of physics selection and filtering algorithms and architectures that are deployed on these commodity systems, it is necessary to investigate the advantages and possible physics impacts of that approach above and beyond the current state-of-the-art. 

\subsection{Fast track reconstruction at the trigger level}
\label{sec:tracking}
The reconstruction of charged-particle trajectories is one of the important experimental tasks in collider physics. Most precision measurements and searches for new physics require reconstruction of all charged-particle trajectories with transverse momentum above a threshold. This task is usually performed with sophisticated software which requires significant computing resources. This approach has been successful but faces challenges on two fronts. Higher data rates will require corresponding increases in resources. Secondly, rare signals involving charged particles may be difficult to identify rapidly if they are not accompanied by other energy deposits in the detectors which can be triggered on. Triggering capability for such events has to rely on fast \textit{track triggers}. 

Algorithms have already been developed to perform rapid reconstruction and momentum estimation of charged-particle trajectories using silicon detectors at small radius in collider experiments~\cite{LOI_IF002, PapertoLOI_IF002}. The speed is achieved by embedding algorithms in highly-parallelized and potentially heterogeneous computing architectures, built using commercial FPGAs and GPUs. Among other benefits, these methods may be used to select rare events containing so-called \textit{disappearing tracks}, i.e. charged particles that decay before reaching the calorimeters. An example of the impact of such signatures is provided by models~\cite{DMseraches_LOI_IF002, NeutralinoDM_LOI_IF002, NeutralinoDM_14_100TeV_LOI_IF002} of particle dark matter whose interaction is mediated by metastable charged particles, such as charginos decaying to neutralino dark matter. 

An expansion of this approach to reconstructing unconventional track signatures for long-lived particles and other exotic processes at high energy colliders is proposed in Ref.~\cite{LOI_EF008}. These new ideas take into account a broader set of challenges to possible triggering strategies: conventional prompt track reconstruction, displaced tracking (high-$d_0$, or impact parameter) and unassociated hit counting.  For example, models such as $R$-parity violating SUSY with displaced leptons, vertices, and jets may offer a variety of processes that would benefit from the capability to identify and select these unconventional track signatures. However, the challenges to doing so are large, and new algorithms leveraging the sophistication allowed by ML approaches and fast heterogeneous processing systems are likely to be crucial in order to realize the potential of these ideas while still meeting the constraints of the system.  

\subsection{Specialized data processing for signal extraction}
\label{sec:signals}
Physics facilities often require specialized data processing solutions to extract the desired signals and to perform real-time analysis which depend on the challenges that those systems face. 
Signal extraction in the low-energy regime proves to be challenging due to the enormous amount of additional, background-like contributions to the high particle density environment. 
High data rates often require a scalable architecture of the DAQ system as well as specialized algorithms that are employed for signal identification and reconstruction.
Computational approaches can present a limiting factor for data analyses due to the need to characterize and analyze an extremely high number of data channels. 
The following are examples to illustrate these challenges.

\subsubsection{Data acquisition and event selection for low-energy events}
\label{sec:dune}
The DUNE Far Detector~\cite{DuneFar} will be generating an unprecedented amount of data, in the form of a continuous, unbiased, and high-resolution ``video'' of ionization charge and scintillation light depositions within the detector’s ultimate 40 kton-fiducial-mass Liquid Argon (LAr) volume. 
The current DUNE Far Detector is designed to target galactic supernova bursts, with current standard reconstruction tools resulting in a visible energy threshold of several MeV. The capabilities for low-energy reconstruction are reasonably well demonstrated with full readout, such as would follow a burst trigger. However more challenging is efficient triggering on individual neutrino events, both for the formation of a burst trigger (which consists of multiple individual events occurring in time coincidence) and for the selection of single, isolated low-energy events. With appropriate detector and data acquisition system developments, much of which are already ongoing, the reach of DUNE can be expanded to provide access to rare and low-energy events down to the $\sim$1\,MeV visible energy range.

Examples~\cite{LOI_IF167} of such developments include a more sophisticated, powerful, and intelligent trigger system, benefiting particularly from recent developments in machine learning and specialized data processing accelerators, used to extract these low-energy signals from abundant radiological backgrounds and detector noise (see, e.g. Refs.~\cite{Dune2, Dune3, Dune4}, and related efforts for other experiments or more broadly, e.g. Refs.~\cite{Dune5, Dune6}).

\subsubsection{Scalable DAQ system for high data rates}
\label{sec:cres}
High data rates are not only a challenge for high-energy colliders and accelerator-based neutrino beams, but also for other facilities like the Project 8~\cite{Project8} experiment that aims to make a direct measurement of the absolute neutrino mass using Cyclotron Radiation Emission Spectroscopy (CRES) and the tritium beta-decay technique. 
In Phase-III of this project, the tritium source will be surrounded by a cylindrical array of 75 microwave antennas to detect the cyclotron radiation from the beta-decay electrons and antenna signals will be combined using digital beam-forming. Those signals from the individual electron events will be detected and reconstructed in real-time by the DAQ system. 

In Ref.~\cite{LOI_IF046} a scalable architecture of the DAQ system, and the advanced algorithms that will be used to identify and reconstruct electron signals is described. The DAQ system will comprise five stages (digitizer, front-end, trigger, tracking and data reduction) with the main challenge being the data rate that will be present all the way through to the tracking stage as both trigger and tracking require all of the data from all channels for a given period of time. Parallization of this process by antenna channel or the physical location of a beam-formed channel is not possible.
To combat these challenges, three guidelines are being followed in the initial design process: 
\begin{itemize}
    \item Data will be moved as little as possible within the system;
    \item Data will be time-multiplexed so that each compute node is considering a single chunk of time;
    \item As many as possible trigger/reconstruction/analysis algorithms and optimized compute hardware based on the algorithm(s) will be explored.  
\end{itemize}

\subsubsection{Multi-messenger astrophysics and gravitational wave signal extraction}
\label{sec:ligo}

At multi-messenger astrophysics (MMA) facilities, astrophysical events such as the binary black hole mergers that create gravitational waves are detected by one or more observatories using triggered data. 
These triggers are then used to alert other telescopes and observatories that an event of shared interest may have been detected and to initiate complementary data collection for potential confirmation and follow-up study of detected events. While these and other systems have filters in place for these operations, they are typically simple linear filters designed based on models of how data corresponding to multi-messenger events would appear. 
Two of the related challenges faced by these facilities are the speed and fidelity with which these astrophysical phenomena can be identified and then communicated in real-time.

Current computational approaches limit gravitational-wave astronomy and LIGO~\cite{LIGO} data analysis in particular. 
Challenges include characterizing thousands of channels of data to provide low-latency data quality evaluation, optimizing interpolation across gravitational-wave waveforms in a 16-dimensional parameter space (including masses, spins, distance, inclination, and sky location), fitting data in this high-dimensional space, and addressing essential scientific questions such as inferring the equation-of-state at supranuclear densities using observations from binary neutron stars and characterizing fundamental tests of Einstein’s theory of general relativity \cite{2017PhRvD96l3011D,2017PhRvD95f2003A,2019arXiv191009740E,2019arXiv191104424D,2018JCAP07048P,2019PhRvL.123a1102A}.
These challenges are currently a computationally limiting factor for low latency estimates of gravitational wave signal significance to enable electromagnetic follow-up of gravitational-wave sources. 

There is thus a significant opportunity to deploy modern TDAQ processing pipelines, potentially including also execution of real-time ML methods, to design, deploy, and commission new trigger or alert systems that (a) {satisfy hard real-time or low-latency processing constraints}; (b) extract features that {jointly leverage heterogeneous, multi-modal data from multiple sources} despite communication constraints; (c) are accompanied by {meaningful confidence intervals}; and (d) allow for {novel triggers or event detectors} based on anomaly detection and semi- or self-supervised learning paradigms.

\section{Opportunities for heterogeneous and dynamic systems}
\label{sec:heterogeneous}
Large-scale physics facilities face challenging demands for high-performance and high-throughput computing resources given the large data sets and high data acquisition rates. As experiments constantly evolve with more sophisticated detector technology and particle beam intensity, so does the complexity of the algorithms that process the ever growing data set.
Making use of \textit{concurrency}, putting multiple processor cores to work on the same problem, is one way to take advantage of the continued increases in computing capabilities. 
In recent years the importance of non-CPU-based computing has significantly increased, despite the number of transistors in a typical processor continuing to roughly double every two years (Moore's Law~\cite{MooresLaw}). The performance of a single core no longer follows this trend, mainly due to thermal constraints in the hardware. 
To improve the computing power, dedicated hardware, e.g. GPUs, FPGAs and ASICs have been developed to accelerate certain kinds of computation by concurrently processing hundreds of thousands of threads. 
Heterogeneous computing, the combination of such a specialized piece of hardware with e.g. a traditional CPU, helps to address the needs of both high-throughput and high-performance compute power, including real-time applications such as those found in current and future TDAQ systems.  
In addition, machine learning is well-suited to these new computing architectures, and many data processing, filtering, and classification algorithms are being rewritten to take advantage of these new developments and opportunities in heterogeneous systems.  


\subsection{On-detector inference and self-driving physics facilities}
\label{sec:selfdriving}
Inference for ``on-detector'' processing with mixed architecture chips is extremely powerful and broadly applicable for any high-speed processing tasks across the intensity, cosmic, energy frontiers of physics, as well as basic energy sciences, nuclear physics, and more~\cite{FastMLWhitePaper,LOI_IF132-2}.  

Amongst the many possible applications at future physics facilities, there are specific proposals to design, construct, and operate integrated, heterogeneous device platforms that implement on-detector inference for real-time data processing and filtering~\cite{FastMLWhitePaper,LOI_IF132}. By moving sophisticated processing closer to the signal collection significant improvements in the data processing pipelines are possible. However, by doing so, the space, power, processing, and even radiation-hardness of the deployed devices is different from that of typical off-detector computing facilities. Consequently, the proposed on-detector capabilities would likely be fulfilled using a mixture of traditional CPUs, Multi-Processor System-On-Chip technology (MPSoC), FPGAs, as well as GPUs, in order to process and adapt to detector and environmental conditions. Recent advances in both the \textit{algorithm} architectures that enable high-performance machine learning (see also \Secref{FPGA_AI}) as well as the \textit{hardware} architectures that can perform these tasks provide a unique opportunity at the intersection of particle physics, computer science, and electronics engineering to develop systems that can process input data, for example from the detectors at the future High-Luminosity Large Hadron Collider (HL-LHC)~\cite{LHC}, more efficiently and with greater accuracy than current approaches.

These concepts often propose to develop sophisticated and hard real-time algorithms for processing high-rate signals using high-performance FPGA/MPSoC architectures including real-time adaptation and dynamic model implementation. As a concrete example, it has been proposed~\cite{LOI_IF132,LOI_IF121} to implement novel physics-inspired neural network image recognition techniques for differentiating signals, for example of Higgs bosons, from the background physics processes using MPSoC and FPGA devices. 

In addition, the concept of a \textit{self-driving trigger system} has been proposed~\cite{LOI_IF072} that is able to learn the hyper-dimensional space of data that are processed -- and potentially discarded -- and thereby autonomously and continuously learn to more efficiently and effectively select, filter, and process data from collected by a detector system. This concept has the potential not only to increase the performance of such systems, but also to increase discovery potential by moving beyond previous paradigms of fixed menus of carefully hand-curated data. The idea to automatically design and refine the trigger and data filtering algorithms at future physics facilities rests heavily on recent advances in explainable architectures~\cite{tibshirani1996regression,letham2015interpretable,ribeiro2016should,macaodha18teaching}, active learning~\cite{tong2001support,liu2004active,chu2011unbiased, bouguelia2013stream,loy2012stream,chen13near,chen15mis}, reinforcement learning~\cite{szepesvari2009algorithms,ghavamzadeh2015bayesian}, and other approaches that take into account the vast availability of simulated and real data, along with the traditional approach to producing a hand-designed trigger menu. The long-term goal of this effort would be to fully exploit the potential of high throughput data steams, and to develop a learning-based framework for stream-based active learning, which takes historical trigger decision records as training data, and learns an efficient active sampling policy on unseen dataset. 

\subsection{FPGA based artificial intelligence inference in triggered detectors}
\label{sec:FPGA_AI}  
In a typical physics detector, each readout channel contains electronics which hosts a local ring buffer where the raw data is stored temporarily while the central trigger decision is generated. At the same time the electronics may provide \textit{trigger primitives} which identify interesting features in the raw data possibly indicating the presence of an important physics event to the trigger node. The logic which generates these trigger primitives must be fast and the resulting stream of data must be small compared to the incoming data bandwidth. This first level data processing logic is a prime candidate for distributed edge AI engines implemented either in the front end ASIC hardware or in FPGAs closely connected to the ASIC or Analog to Digital Converter (ADC).
The amount of time, or latency, allowed for a central trigger decision is dependent on the size of the ring buffer and therefore the time before the data is overwritten. This latency requirement can greatly limit its complexity which may result in excessive false positive triggers. The typical latency requirements for these trigger systems range from 10 $\mathrm{\mu s}$ - 100 $\mathrm{\mu s}$ depending on the type of detector, the raw data rate, and technologies used in the front end. This limit makes software and GPU based AI engines infeasible, requiring the use of ASIC and FPGA based trigger processing systems. FPGAs are the more preferred choice due to their flexibility and lower non-recurring engineering costs. The nature of the FPGA hardware makes it possible to have multiple calculations run in parallel in a pipeline. In other words, the data arrives at the input layer where each node can calculate its result independently. The results are forwarded to the next layer where more sophisticated calculations can be performed, and this ganging can be repeated as necessary to build sufficiently complex triggering decisions. Once the detector is running at full rate, all layers will be calculating simultaneously. This pipelined approach supports both high frame rates and low latency applications.
The intent to continue development of a framework for deploying AI inference engines onto FPGAs is discussed in Ref.~\cite{LOI_IF121}. The idea is to utilize the modern tools provided by the FPGA vendors and the acceleration engines available in modern FPGAs. This framework will allow non professional FPGA developers to define the neural network structure to be deployed on the FPGA as an input to this framework. 

\subsection{FPGA- and GPU-based TDAQ systems}
\label{sec:Mu2e-II}
The LHCb experiment is well-advanced in the development and deployment of a fully GPU-based implementation of the first level trigger~\cite{LHCbCollaboration:2717938,Aaij:2019zbu}.
Mu2e-II~\cite{Mu2e-II}, an evolution of the current Mu2e experiment, will face an increase of the total data rate by a factor of more than $10$. The design of Mu2e-II is not yet finalized, though assumptions are made on the requirements for the TDAQ system to adopt a similar experimental setup as Mu2e, but with improved detector granularity by a factor of 2. The direct consequences of these assumptions are (a) {increased event data size by a factor of 6 to about 1 MB/event}; (b) {reduced period when no beam is delivered (in Mu2e this is 1\,s out of 1.4\,s}; (c) {a factor of 10 larger dose on the electronics}. 
Assuming also that the tape storage capacity will increase by a factor of 2, the trigger rejection factor needs to be about 5 times better than in Mu2e~\cite{Mu2e}. 
Various designs of the trigger and data acquisition system are being considered and are detailed in the respective references: a TDAQ based on GPU co-processor~\cite{LOI_RF041}, a 2-level TDAQ system based on FPGA pre-filtering~\cite{LOI_RF039}, a 2-level TDAQ system based on FPGA pre-processing and trigger primitives~\cite{LOI_RF044}.


\section{Hyper-dimensional data acquisition and processing systems at scale}
\label{sec:systems}
Some of the challenges that are likely to be faced at future physics facilities may require highly specialized solutions that may or may not exist for more common industrial applications. 
For example, extremely high rate applications and high bandwidth requirements, highly tailored or varying temporal data structures, data integrity and robustness, as well as reproducibility requirements for scientific results.
To overcome these and other challenges, it is necessary to further push the envelope of industry standard devices and systems and their intended goals in order to confront and challenge the commodity solutions with the specific needs of low-latency, high throughput and high performance physics facilities. 
TDAQ systems will need to take advantage of new technologies and opportunities in new ways. 
In the following subsections, examples of such paradigm shifting approaches are discussed, along with the innovations required for their realization.

\subsection{Asynchronous first level trigger systems}
\label{sec:asynchronous}
The challenges of distributing and synchronizing stable, low-jitter, high-frequency clocks over a very large distributed system comprised of thousands of optical links will increase in order to address the needs of the experiments at future colliders. Moreover, the precision of the timing to be maintained will reach the level of tens of picoseconds, and the number of channels and processing boards will greatly increase with the increasing granularity of the experiments. Rather than maintain a synchronous system for the full data path to reach a first level trigger decision, it is proposed to tag the data with a time marker only at the very front-end of the detector electronics and transmit and process the data subsequently asynchronously as is already done traditionally for the data acquisition and high-level triggers after this first level. Effectively, the event builder infrastructure moves to process data at the full rate from the detectors. Already some portions of the first level trigger systems at the LHC experiments run asynchronously: the optical data links connecting the boards comprising the trigger system, for example, do not run synchronously since the frequency of the data link operation does not always match a multiple of the machine frequency. Additionally, some experiments also have introduced a “time (de)multiplexing” approach to serve a complete set of detector data from a particular beam crossing to an individual board of the first level trigger. This proposal just extends this time-multiplexing concept to a fully asynchronous event builder architecture.
% ...

Additional benefits of an asynchronous system include a blurring of the lines between the first level trigger, which typically runs in fast FPGAs, and the higher levels, which typically run on CPUs and now also GPUs. In some sense, a set of FPGA, GPU, and CPU processors can be used to execute a mix of traditionally very fast algorithms for the initial selection of data and more complex algorithms that typically take more time for the final selection.

A small-scale prototype of such an asynchronous trigger system for the formation of track from the data of several muon detectors was developed and tested in Ref.~\cite{4179279}. The front-end trigger electronics of three spare cathode-strip chambers of the CMS Endcap Muon system were upgraded to perform pattern recognition and bunch-crossing assignment from the anode data at an 80 MHz frequency. Trigger primitives from up to 3 chambers were transmitted via 10 Gb/s serial links to a newly designed track-finding processor, a PC plug-in card, that could generate the asynchronous trigger acceptance signal with a time marker that is sent back to front-end boards for data read-out.

\subsection{Edge computing devices for detectors}
New experiments are producing increasingly growing amounts of data that, in part, need to be stored, but also need to be processed on the fly, helping in critical decisions in the operation of experiments, such as for example triggering. Individual streams of data carry majorly ‘zeros’ and noise and the whole activity context is not known. The bandwidths of links cannot cope with streams of produced unprocessed data. Also sending of data entails significant energy expense that can be reduced if data is converted to information as early as possible. The streams start reflecting zoned activities in the experiment only when combined in data concentrators. Traditionally, data is minimally processed inside and processing needing context is outside real time or offline. In the regard of the above, it is worth considering techniques of reduction of volumes of raw data generated at detector frontends, and where streams of data get aggregated by embedding machine learning, often termed edge computing in custom designed detector readout electronics. Introduction of smartness by bringing processing inside with artificial neural networks is a future for Read Out Integrated Circuits (ROIC) ASICs. 

The approach pursued is to study and carry out the development of conventional Von-Neumann and non Von-Neumann neuromorphic computing based AI ASICs for scientific data processing being targeted for the front-end electronics, respecting specific needs, such as extreme environments of operation, restriction on power dissipation or circuital resources. This approach postulates harnessing the co-designing methodology. More details on this approach can be found in~\cite{LOI_IF180}. As architectures and sizes of a neural processor fits only classes of problems, first, building and optimizing a neural network model with the tools available today: Tensor Flow, PyTorch or Caffe2 frameworks needs to be done. Then, through training of the neural network model to estimate kernel weights transition can be fed to hardware design. There, hardware constrained high-level synthesis to optimize the registry-transfer-level coding shows up as the right way for implementation due to large sizes of the resulting circuits. These steps need collaboration of experimental physicists, computer scientists and circuit designers to well understand and to work out practices that will be suited optimally for building neural processors for scientific data.

At present, High Energy Physics (HEP) experiments develop machine learning analysis software and GPUs, Digital Signal Processors (DSPs) or FPGAs as hardware allowing re-programmability and versatility. Experiments such as Minerva, DUNE or experiments on the LHC use neural networks for data analysis~\cite{Edge1, Edge2, Edge3}. Until today, however, Neuromorphic Computing algorithms, requiring non Von-Neumann hardware development have been in early stage, and have realistically been not used for any high-scale application. Neuromorphic computing based on GPUs or FPGAs has huge latency and is definitively not a power efficient solution as it involves an enormous number of read and write operations between memory and computation units.

\subsection{Streaming DAQ}

Traditional DAQ systems, and many operating today from the smallest test setups to the most complex at our major facilities, share a common architectural concept of having pipelined and triggered detector readout. This is borne out of the necessity to overcome bandwidth and rate limitations of the available electronics, computing, networking and storage. The assumption is that it would be impossible to read out detectors at the full rate, and even if that were possible, one could not process the data at that full rate, and even if that could be done, that this would create prohibitively large volumes of data that would pose an unsustainable burden on offline computing. However, continuing advances in computing, networking and storage technologies have begun to challenge these long held assumptions. 

Pioneering work over the last decade, in particular by the LHCb experiment
(see for example~\cite{LHCbCollaboration:2014vzo,LHCbCollaboration:2717938,Aaij:2019zbu}),
has pointed the way towards a streaming DAQ mode, in which data are read out as a continuous stream, tagged with timing information that allows the data to be reconstructed when they are at rest. As this is becoming feasible, the concept is also emerging elsewhere, for instance in the Nuclear Physics community (see for example~\cite{JLab:StreamingDAQ}). It is bound to gain further momentum in our field as streaming DAQ offers several advantages over the pipelined trigger mode.

In triggered DAQ, analogue signals from the detector are digitized at a fixed clock rate and written to memory, typically organized as circular buffers. A trigger system reads all or part of these data, often over separate data links, and processes them to make a decision on whether to keep or discard the content of the buffer. When the buffer is to be kept, data readout is initiated and the data are passed over the main DAQ path on to the next stage. The latency of the trigger decision is afforded by the depth of the pipeline and allows for the processing required to reduce the data rate on the fly. The trigger system can use inputs from several detector components and become rather complex in order to achieve this goal. The readout frames are labeled with which trigger they belong to but their readout occurs randomly due to the latency of the trigger decision. Downstream event building then has to collect and arrange all contributions while the data are in motion. This scheme, while hugely successful, can have a number of known limitations. The trigger decisions introduce selection biases, they don't deal well with event pileup, and their complexity is constrained by the front-end electronics and the need to execute within the deadtime path. And the data reduction achieved this way generally comes at a loss of information.

As an alternative to this approach, streaming DAQ aims to read out many parallel continuous streams of data that are encoded with their original time and detector location. The flow of data can be reduced (often greatly) by applying thresholds and zero suppression locally for each stream. Multiple streams can be aggregated into a single hardware link or transport channel. This is referred to as \textit{stream aggregation}. The resulting reductions can be substantial because detectors with a (very) high number of channels typically have (very) low occupancy. Moreover, this can be implemented with generic components because the aggregation is time-based and thus agnostic to the payload of the data. The aggregation of streams also provides an opportunity for further in-line processing, for instance, clusters can be formed from hits. This step is referred to as \textit{stream translation}, where data of one type are received and data of a different type are sent out. Event building now takes place when the data are at rest. Further processing and filtering can be done asynchronously, after the data have been acquired, and the time this can take is constrained by the depth of the intermediate storage rather than the front-end buffers. The temporary storage of the raw data also provides opportunities for the experiment to process the unbiased data stream and to preserve portions of it for studies without having to reconfigure the system. It also gives rise to raw data samples for calibrations or monitoring without the need for special runs. From an architectural perspective, streaming DAQ can be viewed as leveraging higher performance and more cost-effective computing for a much simpler DAQ design, where that is possible, without a complex custom hardware-based trigger, with no independent trigger path but a single DAQ path from the detector, and with a deterministic time-ordered data transport system with tiered storage that can provide opportunities for more streamlined and sequential workflows.

\section{Installation, commissioning, integration, and operations}
\label{sec:operation}
To bring the innovative detector-related research and development to life and to perform discovery science with next-generation facilities, equal attention must be paid to -- and adequate effort be invested in -- their installation, commissioning, integration, and operation. Without these crucial efforts, data-taking will be compromised and science goals will not be met. Experience from previous deployments in large physics facilities has shown that a lack of attention in this area can lead to substantial and costly delays, failure to reach design goals, as well as insufficient performance, all jeopardizing the physics output of the experiment. 
It is therefore vital for the success of innovations that their installation and commissioning is carried out successfully and that a high level of operational efficiency is achieved and maintained.

New components have to undergo a transition from final testing to their installation, commissioning, and integration, which brings along new challenges beyond those encountered during construction. 
Many of today's physics facilities will be upgraded in the future with new detector and instrumentation capabilities, which bears the additional challenge of getting the new components to work in concert with the already existing ones.

Operation of small, medium, and large-scale physics facilities over long periods of time is a major challenge that needs to remain a high priority during the data-taking period to ensure successful and efficient collection of high quality data.
Over time, operations are often streamlined with the benefit of experience, and operational efficiency is improved through diligent effort. Nonetheless, these incremental improvements may not be sufficient to overcome the operational challenges of the loss of both expert personnel and knowledge over long periods of time.

High-efficiency operations demand a substantial level of effort that needs to be maintained as long as the experiment is running. 
The science is only made possible by operating these extensive and sophisticated systems, which have been built to collect unique and novel datasets, to their full potential.

\comment{
    \begin{itemize}
        \item What foundational and operational skills and training do we need to be able to achieve the goals?
        \item ? soft skills: communication, discussion, presenting a personal point of view and working in a group
        \item What are the interfaces to other areas, fields etc? 
        \item Acquisition of domain knowledge and sharing of such technical and domain knowledge
        \item Recruit talent to do the actual work
        \item Expertise that we need to build
        \item Connect to the design, building, and execution
        \item Need pool of people who actually know about devices, etc.
        \item Workforce development 
\end{itemize}
}

\section{Building and retaining domain knowledge and technical expertise}
\label{sec:people}

Realizing the scientific potential of these facilities rests on the development and deployment of innovative technologies. Consequently, realizing those technologies requires building and retaining the relevant domain knowledge and technical expertise. As such, it is critical for our field to recognize the importance of the highly skilled people that are carrying out this work. 

Without the foundational and operational skills contributed by technicians, engineers, graduates, doctoral students, postdoctoral researchers, research fellows, university academics and laboratory staff, it is not possible to deliver discovery science at physics facilities. 
However, the availability and recruitment of staff with the necessary skills and research interests for the specialized needs of detector installation, commissioning, integration and operation is problematic as career progression and promotion is challenging for these highly skilled people. In case of a successful hire, retention is an issue. Technical staff and engineers move to industry for higher pay. The availability of positions that allows early career scientists to stay on the path of a career in instrumentation is limited. Graduates, postdoctoral researchers or research fellows often move on to physics analyses as they have difficulties to secure permanent academic positions working in detector-related areas.
These factors make it extremely difficult to build a community of experts who are able to eventually operate the detector efficiently for years to come. 
Scientific progress rests on current and future physics facilities that have diverse technical needs. Therefore it is in the vital interest of our scientific community to match that need with career opportunities for people to build the domain knowledge and technical expertise. 

\subsection{Gaining expertise: new hardware, firmware and software}
The expertise necessary to realize the scientific potential spans the areas of hardware, firmware, and software development, and requires a diverse and specialized set of skills.
Many of these required skills are due to the innovation of new processors and accelerator architectures, with one aspect of the diversity arising from e.g. being able to create AI algorithms and to make them run efficiently on FPGAs (see for example \Secref{FPGA_AI}).
Industry toolchains will need to be adapted and integrated into the workflows defined by the innovations in future physics facilities. 
While new skill sets do need to be acquired by our community, there is also the need to extend the existing domain knowledge into the area of the new technologies in order to evolve and enhance the existing knowledge in this new area for maximum efficiency and performance. 

\subsection{Operational experience}

\begin{quote}
\textit{''[Shifters and experts] wanted for hazardous journey. Small wages, bitter cold, long months of complete darkness, constant danger, safe return doubtful. Honour and recognition in case of success.``}
\begin{flushright}-- Alleged advertisement by Sir Ernest Shackleton for the voyage of the \textit{Endurance} to Antarctica
\end{flushright}
\end{quote}
In addition to the more general aspects of technical expertise and skills, operating a detector or part of a detector requires intimate knowledge of particular technical implementations and of the facilities themselves.
This is something that can only be obtained in an optimal way by developing familiarity within the facilities. The knowledge about them has to be transferred within the facilities.
Gaining operational experience is achieved through employment in a position that includes the responsibility for the operational control of all or part of this facility, and it is very much on-the-job training.

\comment{
\begin{itemize}
    \item  Coordination with Community Engagement Frontier, Topical Group CommF1: Applications and Industry 
    \begin{itemize}
        \item Bruhwiler et al, ``Collaboration between industry and the HEP community`` \href{https://www.snowmass21.org/docs/files/summaries/CommF/SNOWMASS21-CommF1_CommF0-AF0_AF1_Bruhwiler-066.pdf}{LOI 66}
        \item Mase and Kikuchi, ``Technology transfer from KEK to industry: What is needed to commercialize the technology developed in high energy`` \href{https://www.snowmass21.org/docs/files/summaries/CommF/SNOWMASS21-CommF1_CommF0_Kazuhiko_Mase-024.pdf}{LOI 24}
    \end{itemize}
\end{itemize}
}

\section{Conclusions and Outlook}
\label{sec:conclusions}
The trigger and data acquisition systems of future physics facilities are facing a variety of challenges that need to be addressed by innovations, a few examples of which of either existing or proposed solutions are given in this White Paper. 
We are considering three overarching themes: novel applications of current technologies, adapting technology drivers to specialized needs, and building domain knowledge and community effort.
In each of these areas, important questions remain to be answered that will lead our field in moving forward.

Designing and building large-scale, heterogeneous, and dynamic systems for data acquisition, filtering, processing, and storage can make use of both commodity and custom hardware, firmware, and software.
    \begin{itemize}
        \item What is the current state of the art for heterogeneous digital and analog solutions and how can we benefit from this?
        \item What physics selection and filtering algorithms and architectures can be deployed on these commodity systems?
        \item What are the advantages and potential physics impacts of that approach above and beyond the current state of the art?
    \end{itemize}

Our community will have to consider and explore paradigm shifting approaches that can take TDAQ systems to new levels by taking advantage of new technologies and opportunities in new ways. Pushing the envelope of industry standard devices and systems, and their intended goals, by confronting these commodity solutions with the specific and specialized needs of low-latency, high throughput and high performance physics facilities is key. 
    \begin{itemize}
        \item How can we take devices, firmware, and software designed by industry for industrial applications and build the systems we need to do physics with? What innovations are needed to do this?
        \item What are the particular challenges that we face that are niche or nonexistent in industrial applications? 
        \item What can the scientific community feed back into industry in order to adapt technologies and/or create new ones that meet the goals and challenges of next-generation physics facilities?
        \item How do the new technologies influence our architectural concepts?
    \end{itemize}

Installation, commissioning, integration and operation are vital phases on the way of realising any future facility. They demand an effort by the community and its people that involves identifying, acquiring and retaining the needed domain knowledge, technical expertise, skills and experience in the wide landscape of new technologies. 
    \begin{itemize}
        \item What foundational and operational skills and training do we need in our community to successfully build and operate the future facilities?  
        \item What is the expertise that we need to gain and foster? How can domain knowledge be acquired and retained?
        \item How can we recruit talent to do the work? How can we maintain a pool of experts with knowledge about the system specifics?
        \item How can operations be tied to the construction of the system? How can we enable knowledge transfer along the different phases of the experiments? How can the operational experience inform and influence the design of the future systems?
    \end{itemize}

\clearpage
\bibliographystyle{JHEP-mod}
\bibliography{references} 


\end{document}